\documentclass[a4paper]{jpconf}
\usepackage{graphicx}
\begin{document}
\title{Reaction Plane and Beam Energy Dependence Of The Balance Function at RHIC}

\author{Hui Wang for the STAR collaboration}

\address{W216, National Superconducting Cyclotron Laboratory, Michigan State University, East Lansing, MI, 48824-1321, USA}

\ead{wang@nscl.msu.edu}

\begin{abstract}
The balance function, which measures the correlation between opposite sign charge pairs, is sensitive to the mechanisms of charge formation and the subsequent relative diffusion of the balancing charges. The study of the balance function can provide information about charge creation time as well as the subsequent collective behavior of particles. In this paper, we present a reaction-plane-dependent balance function study for Au+Au collisions at $\sqrt{s_{\rm NN}}$ = 200 GeV and compare with results from recent three particle correlation measurements. We also report balance functions for relative pseudorapidity ($\Delta \eta$), relative rapidity ($\Delta y$), and relative azimuthal angle ($\Delta \phi$) from the recent RHIC beam energy scan data.
\end{abstract}

\section{Introduction}

The study of correlations and fluctuations can provide information about the properties of the hot and dense matter created in relativistic heavy-ion collisions. one such observable, the balance functions, which measures the correlation between the opposite sign charge pairs, are sensitive to the mechanisms of charge formation and the subsequent relative diffusion of the balancing charges \cite{balance_PRL}. Due to conservation laws like electric charge conservation, particles and their anti-particles are pair produced and correlated initially in coordinate space, if a delayed hadronization occurs, the lower temperature and less expansion and diffusion will result in a narrower charge balance function.

The reaction-plane-dependent balance function in azimuthal angle can be written as

\begin{eqnarray}
\fl
B(\phi ,\Delta \phi ) = \frac{1}{2}\{ \frac{{\Delta _{ +  - } (\phi ,\Delta \phi ) - \Delta _{ +  + } (\phi ,\Delta \phi )}}{{N_ +  (\phi )}} + \frac{{\Delta _{ -  + } (\phi ,\Delta \phi ) - \Delta _{ -  - } (\phi ,\Delta \phi )}}{{N_ -  (\phi )}}\} .
\end{eqnarray}

Here  $N_{+(-)}(\phi)$ is the total number of positive(negative) particles that have an azimuthal angle $\phi$ with respect to the event plane, $\Delta_{+-}(\phi,\Delta \phi)$ represents the total number of pairs summed over all events where the first (positive) particle has an azimuthal angle $\phi$ with respect to event plane and the second (negative) particle has a relative azimuthal angle $\Delta\phi$ with respect to the first particle. Similarly we can express $\Delta_{++}(\phi,\Delta \phi)$, $\Delta_{-+}(\phi,\Delta \phi)$ and $\Delta_{--}(\phi,\Delta \phi)$.

On the other hand, it has been discussed recently that the hot and dense matter created in heavy ion collision may form metastable domains where the parity is locally violated, this possible local parity violation \cite{LPV} coupled with strong magnetic field produced by passing nuclei in such a collision could cause a charge separation across the reaction plane in non-central collisions called the chiral magnetic effect (CME) \cite{CME_1,CME_2,CME_3}. One observable proposed to measure the Chiral Magnetic Effect is the three point correlator \cite{3_point_correlator} ,where $\phi$ is the azimuthal angle of a particle and $\psi _{RP}$ is the reaction plane angle.

\begin{eqnarray}
 \gamma_{\alpha, \beta}=< \cos (\phi _\alpha  + \phi _\beta  - 2\psi _{RP} ) > .
\end{eqnarray}

\section{Experimental Method}

The data used in this analysis is from Au+Au collisions at $\sqrt{s_{\rm NN}}$ = 200 , 62.4, 39, 11.5, and 7.7 GeV taken by the STAR experiment. The Time Projection Chamber (TPC) was used as the main detector 	for charged particle tracking. All tracks were required to have a distance of closest approach (DCA) to the measured event vertex of less than 3 cm. For all charged particles, a transverse momentum cut of $0.2 < p_{t} < 2.0$ GeV/$c$ was applied while for identified particles, we used a $p_{t}$ cut of $0.2 < p_{t} < 0.6$ GeV/$c$.  The pseudorapidity cut used is $|\eta| < 1.0$. Particle identification was done by selecting particles whose specific energy losses($dE/dx$) were within two standard deviations of the energy-loss predictions for a given particle type and momentum. In addition, electrons were excluded from the analysis for all cases by specific energy loss inside the TPC. To correct for differences between the acceptances for positive and negative particles, a mixed event subtraction is used for all balance function results throughout this paper.

To determine the event plane angle, the second Fourier harmonic in the azimuthal angle distribution is used. We used all tracks from the TPC while a $p_{t}$-weight method was used to maximize the event plane resolution. Also the $\phi$ weight method is applied to flatten the event plane distribution.

\section{Results}

\subsection{Event-Plane-Dependent Balance Function at  $\sqrt{s_{\rm NN}}$ = 200 GeV}

\begin{figure}[h]
\begin{minipage}[t]{18pc}
\includegraphics[width=18pc]{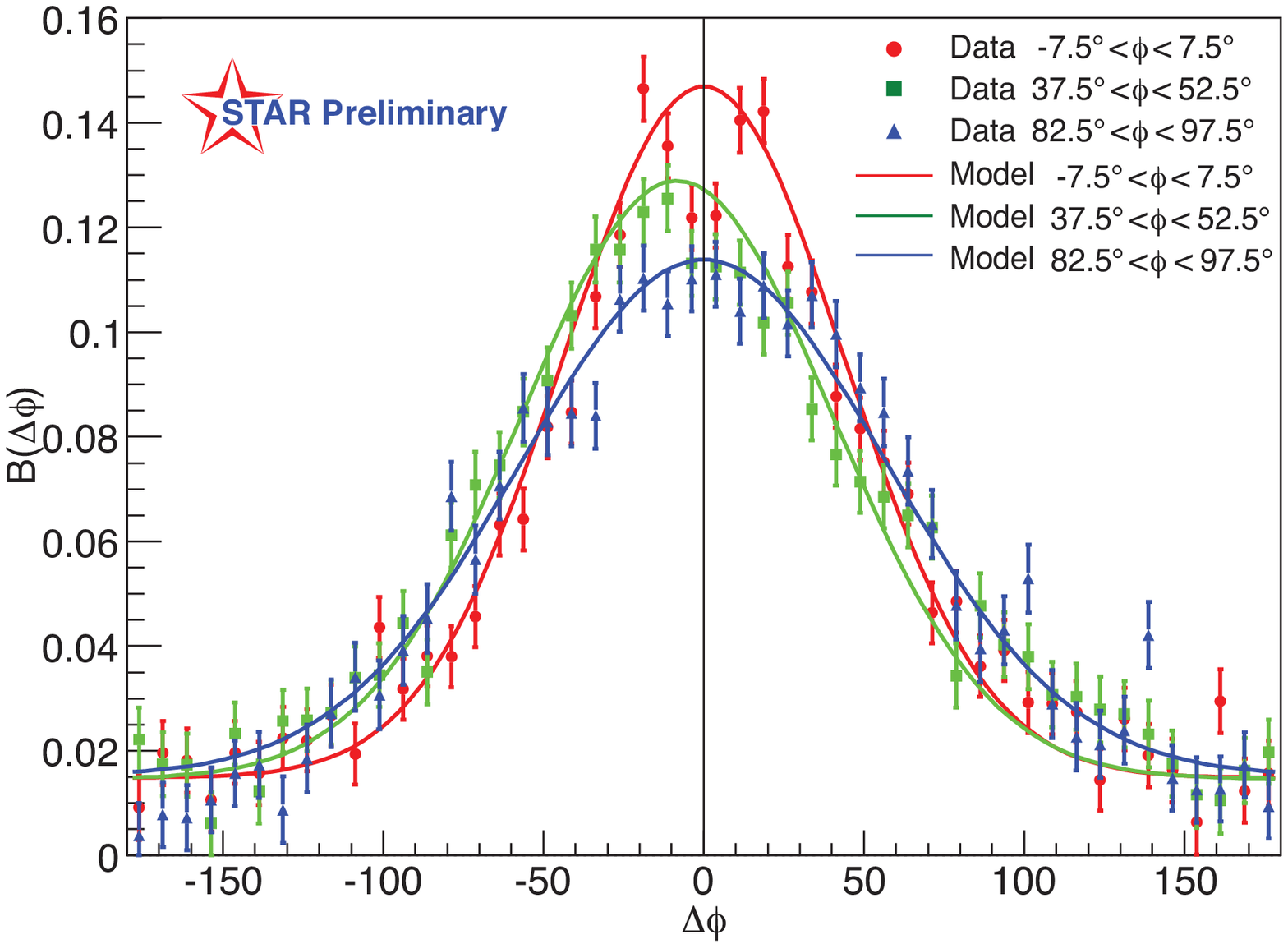}
\caption{\label{fig:fig01}(Color online)  The balance function for $\phi = 0^{\circ}$ (in-plane), $\phi = 45^{\circ}$, and $\phi = 90^{\circ}$ (out-of-plane) particles pairs. The 40-50\% centrality bin is shown.   The points are from the experimental data (not corrected for event plane resolution), while solid lines represent the blast-wave model calculations \cite{parity_soeren}.}
\end{minipage}\hspace{2pc}%
\begin{minipage}[t]{18pc}
\includegraphics[width=18pc]{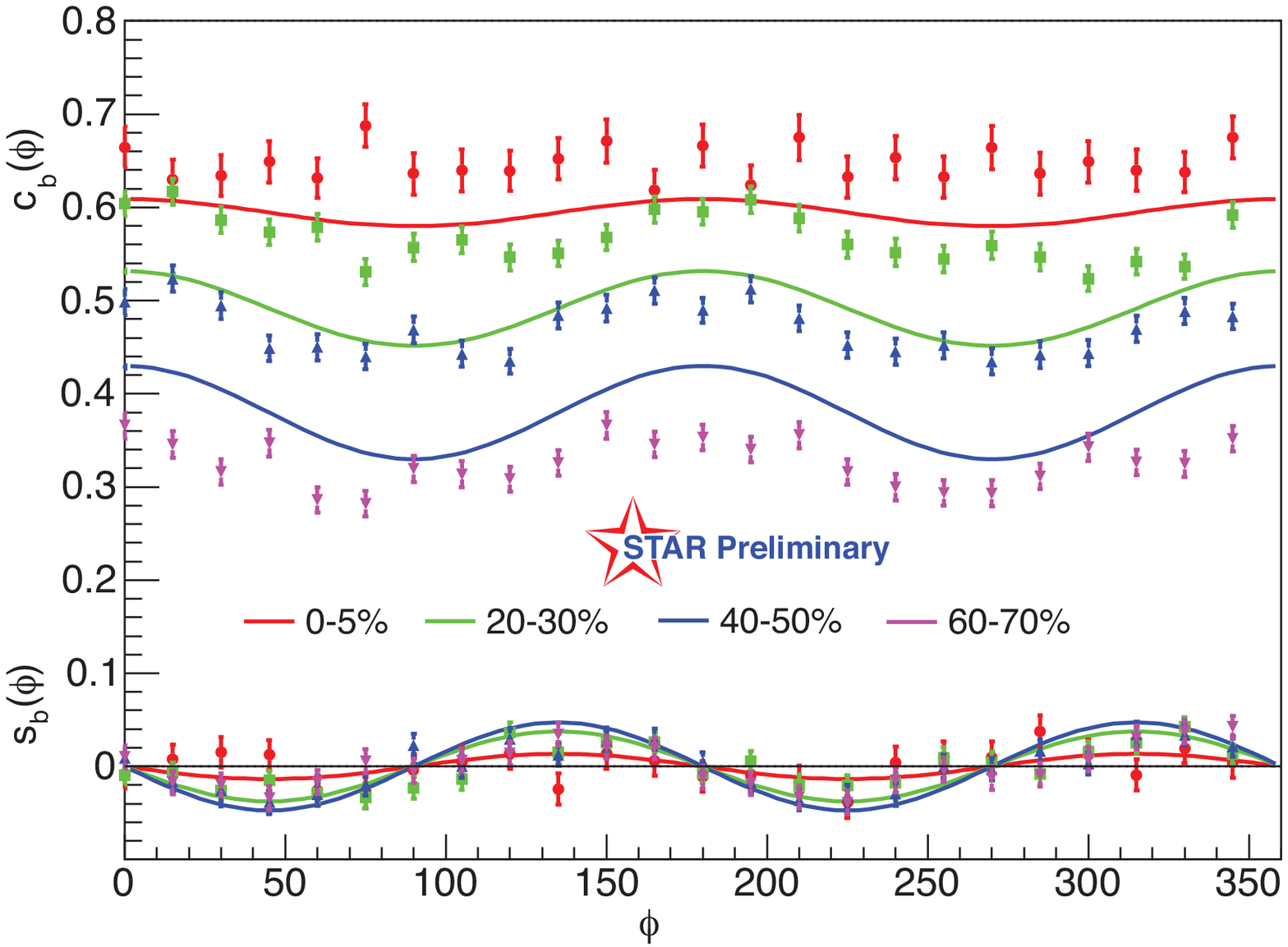}
\caption{\label{fig:fig02}(Color online)  The weighted average cosine and sine of balance function,  four centralities are shown here. The points are from the experimental data (not corrected for event plane resolution), while solid lines are from the blast-wave model \cite{parity_soeren}.}
\end{minipage} 
\end{figure}

Figure~\ref{fig:fig01} shows $\phi = 0^{\circ}$ (in-plane), $\phi = 45^{\circ}$, and $\phi = 90^{\circ}$ (out-of-plane) balance function for 40-50\% centrality only. The in-plane balance function is narrower than the out-of-plane balance function, which is caused by the stronger collective flow in-plane: charge pairs are created closely in space and time due to conservation laws, this correlation would remain to the final stage if they are not diffused, which gives a narrower balance function.  Figure~\ref{fig:fig01} also shows the $\phi = 45^{\circ}$ balance function, which is asymmetric and peaked at negative $\Delta \phi$.  Because of the strong elliptic flow established at non-central collision, if charge pairs are emitted with an angle respect to the reaction plane, the correlation would be stronger on the in-plane side compare to out-of-plane side.  Also shown are the blast-wave model calculations \cite{parity_soeren}, we can see that for all three cases here, the blast-wave model agree well with experimental data.

To quantify the collective flow effect on balance function, we also study the weighted average cosine, $c_b(\phi)$, and sine, $s_b(\phi)$, extracted from the balance functions.

\begin{eqnarray}
\fl
\nonumber
c_b (\phi ) = \frac{1}{{z_b (\phi )}}\int {d\Delta \phi } B(\phi ,\Delta \phi )\cos (\Delta \phi ),\\
\nonumber
s_b (\phi ) = \frac{1}{{z_b (\phi )}}\int {d\Delta \phi } B(\phi ,\Delta \phi )\sin (\Delta \phi ), \\
z_b (\phi ) = \int {d\Delta \phi B(\phi ,\Delta \phi )}.
\end{eqnarray}

$c_b(\phi)$ represents the width of balance function. If charges are created at the same point and did not diffuse due to strong collective flow, $c_b(\phi)$ would be close to unity.  $s_b(\phi)$ is an odd function of $\Delta\phi$, so it can quantify the asymmetry of balance function. Figure~\ref{fig:fig02} shows $c_b(\phi)$ and $s_b(\phi)$ for Au+Au  collisions at  $\sqrt{s_{\rm NN}}$ = 200 GeV. In the figure, $c_b(\phi)$ is closer to unity in the 0-5\% centrality bin, which is due to a stronger collective flow in central collisions, while in mid-peripheral and peripheral collisions,  $c_b(\phi)$ shows a difference between the in-plane and out-of-plane balance functions, which is caused by stronger elliptic flow in-plane. For  $s_b(\phi)$, it reaches maximum at $\phi {\rm{  =  }}135^{\circ},315^{\circ}$ and minimum at $\phi {\rm{  =  }}45^{\circ},225^{\circ}$, which demonstrates that charged pairs are more correlated on the in-plane side when emitted with an angle $\phi$ with respect to the reaction plane.

Figure~\ref{fig:fig02} also shows a comparison with the blast-wave model \cite{parity_soeren}. The blast-wave model includes a breakup temperature $T_{kin}$, the maximum collective velocities in the in-plane and out-of-plane directions, the spatial anisotropy of the elliptic shape by fitting STAR published $v_{2}$ and spectra data \cite{STAR_v2}, this model also assume local charge conservation and initial separation of balancing charges at freeze-out by fitting experimental results \cite{balance_PRC}. The difference between data and the blast-wave model could due to the finite event plane resolution for the data.

The difference between the same-sign and opposite-sign three point correlator $\gamma_{\alpha \beta}$ can be expressed as \cite{parity_soeren} 

\begin{eqnarray}
\fl
\gamma _p  = \frac{1}{2}(2\gamma _{ +  - }  - \gamma _{ +  + }  - \gamma _{ -  - } ) = \frac{2}{M}[v_2  < c_b (\phi ) >  + v_{2c}  - v_{2s} ],
\end{eqnarray}

where
\[
\begin{array}{l}
\fl
 v_{2c}  = < c_b (\phi )\cos (2\phi ) >  - v_2  < c_b (\phi ) >,  \\ 
\fl
 v_{2s}  =  < s_b (\phi )\sin (2\phi ) >,  \\ 
 \end{array}
\]

and the bracket represents
$$ 
\fl
< f(\phi ) >  = \frac{1}{M}\int {d\phi \frac{{dM}}{{d\phi }}z_b (\phi )f(\phi )} . 
$$

In this equation, $v_2 \left\langle {c_b (\phi )} \right\rangle $ will be positive if there are more charge pairs in-plane than out-of-plane, $v_{2c}$ will be positive if the charge pairs are more correlated in-plane than out-of-plane, while $v_{2s}$ will be negative if the charge pairs are more correlated on the in-plane side.

\begin{figure}[h]
\begin{minipage}[t]{18pc}
\includegraphics[width=18pc]{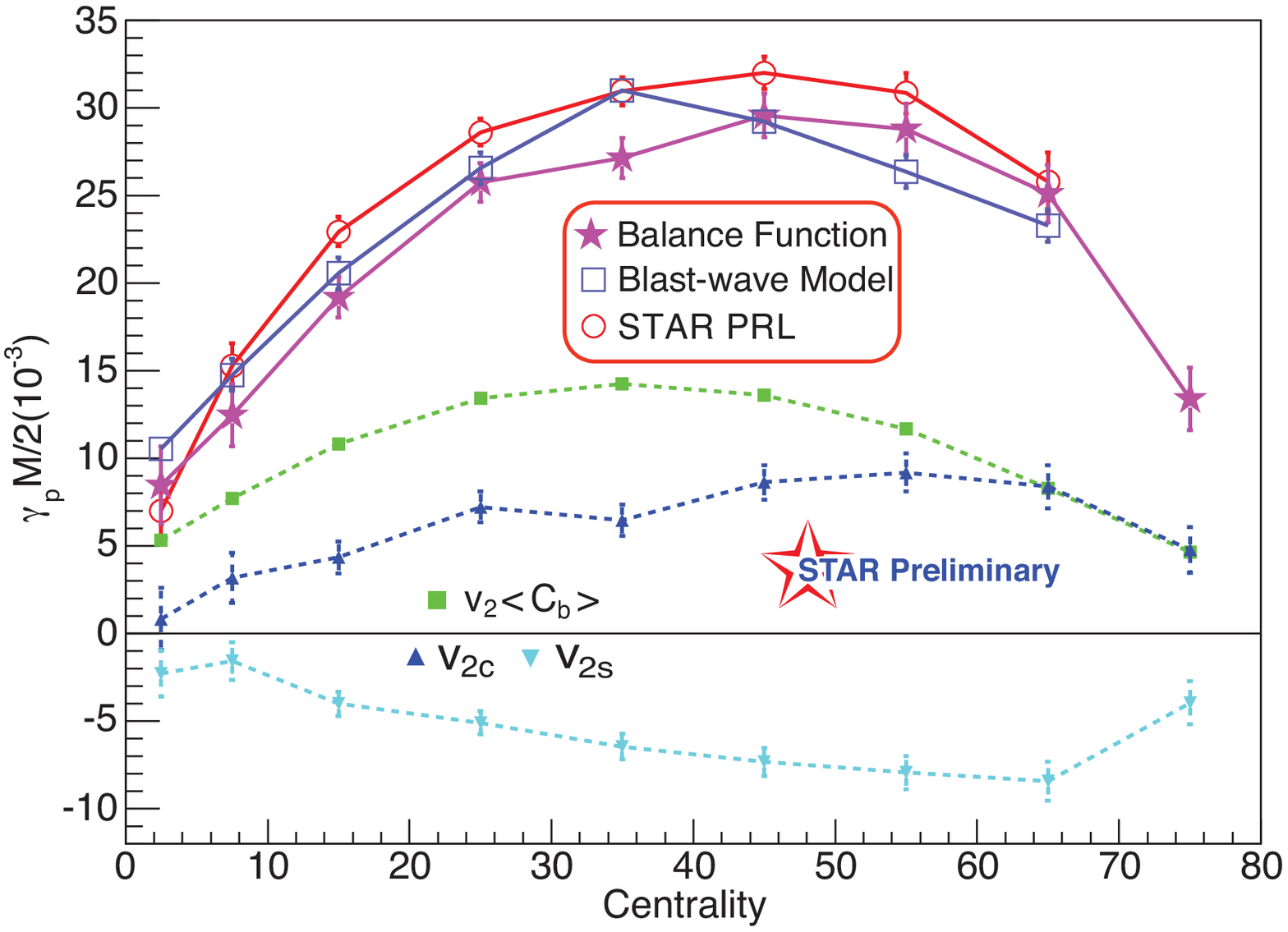}
\caption{\label{fig:fig03}(Color online)  parity observable $\gamma_P$ scaled by experimental multiplicity}
\end{minipage}\hspace{1pc}%
\begin{minipage}[t]{19pc}
\includegraphics[width=19pc]{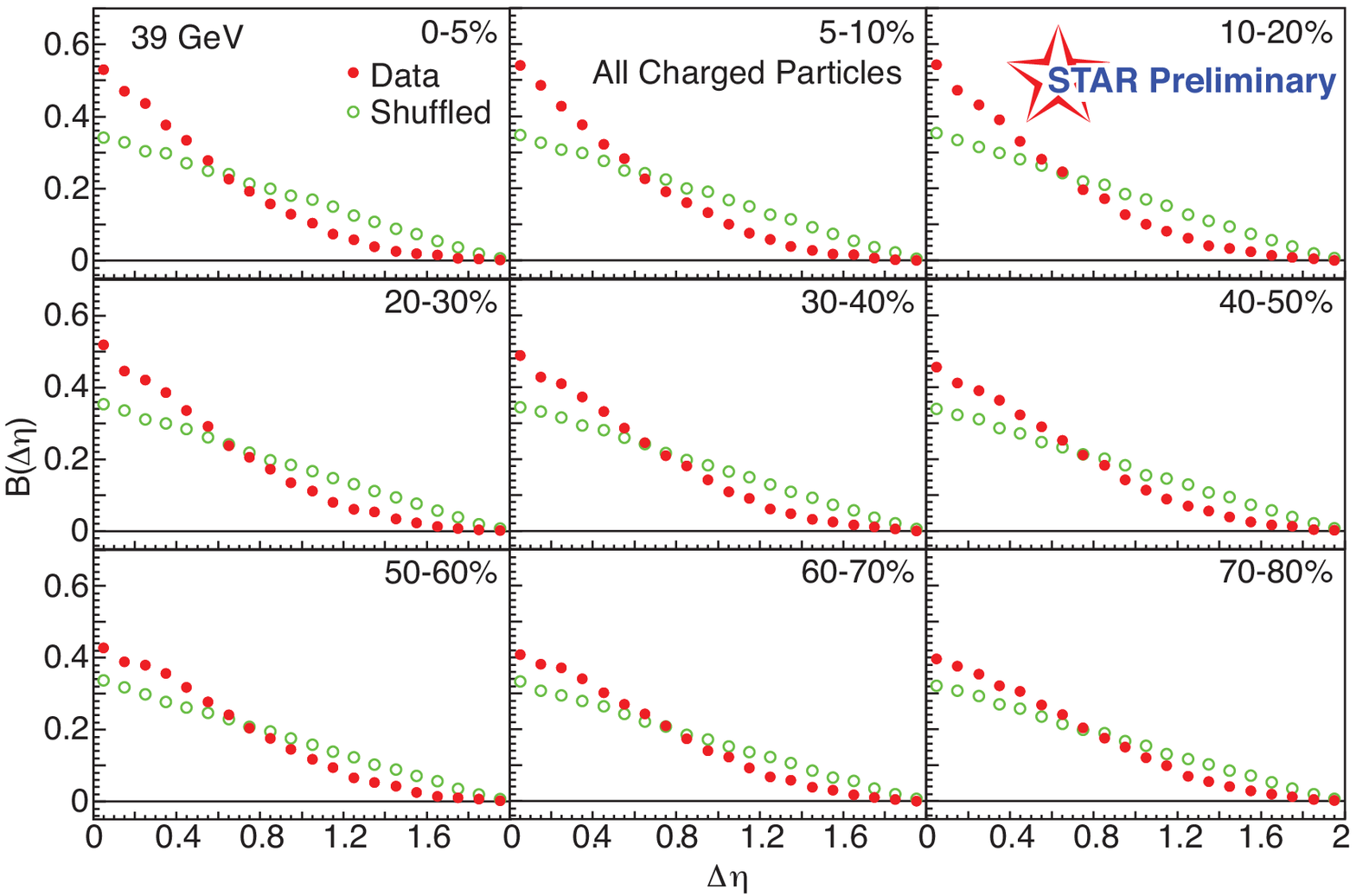}
\caption{\label{fig:fig04}(Color online) The balance function in terms of  $\Delta \eta$ for all charged particle pairs from Au + Au collisions at $\sqrt{s_{\rm NN}}$  = 39 GeV for nine centrality bins.}

\end{minipage}

\end{figure}

Figure~\ref{fig:fig03} shows the parity observable calculate from balance function as well as its three components. All data points are corrected for the event-plane resolution here. To compare with previous results, we also plot the $\gamma_P$  from STAR published data \cite{parity_PRL} scaled by the measured uncorrected multiplicity in the same plot. Mathematically, the balance function result should equal the one from $\gamma_P$ and they do agree well. We can also see that, A thermal blast-wave model \cite{parity_soeren} incorporating local charge conservation and flow reproduces most of the signal.

\subsection{Balance Function at  $\sqrt{s_{\rm NN}}$  = 39 GeV}

Previously, STAR has published measurements of the balance function  from Au + Au, d + Au, and p + p collisions at  $\sqrt{s_{\rm NN}}$  = 200 GeV \cite{balance_PRC}. In that paper, a simplified(Eq.(\ref{eq:balance}), no reaction-plane-dependence) balance function was used to measure the charge pairs correlation in terms of relative pseudorapidity, $\Delta\eta$, relative rapidity, $\Delta y$, relative azimuthal angle, $\Delta\phi$, and invariant relative momentum, $q_{inv}$

\begin{eqnarray}
\label{eq:balance}
\fl
B = \frac{1}{2}\{ \frac{{\Delta _{ +  - }  - \Delta _{ +  + } }}{{N_ +  }} + \frac{{\Delta _{ -  + }  - \Delta _{ -  - } }}{{N_ -  }}\} .
\end{eqnarray}

It has been shown that the balance function in terms of $\Delta \eta$ and $\Delta \phi$ for all charged particles and in terms of $\Delta y$ for charged pions narrow in central Au + Au collisions. Here we perform a similar measurement from Au+Au collisions at $\sqrt{s_{\rm NN}}$  = 39 GeV.

Figure~\ref{fig:fig04} shows the balance function in terms of $\Delta \eta$ for all charged particles for nine centrality bins. The balance function for central collisions is much more narrower compare to the one from peripheral collisions. The balance function for shuffled events is much wider than the balance functions from the measured data. The shuffled events are created by randomly shuffling the charges of the particles in each event. By doing this, all charge-momentum correlations are removed in shuffled events, which could produce the widest balance function within the same experimental acceptance. 

\begin{figure}[h]
\begin{minipage}[t]{18pc}
\includegraphics[width=18pc]{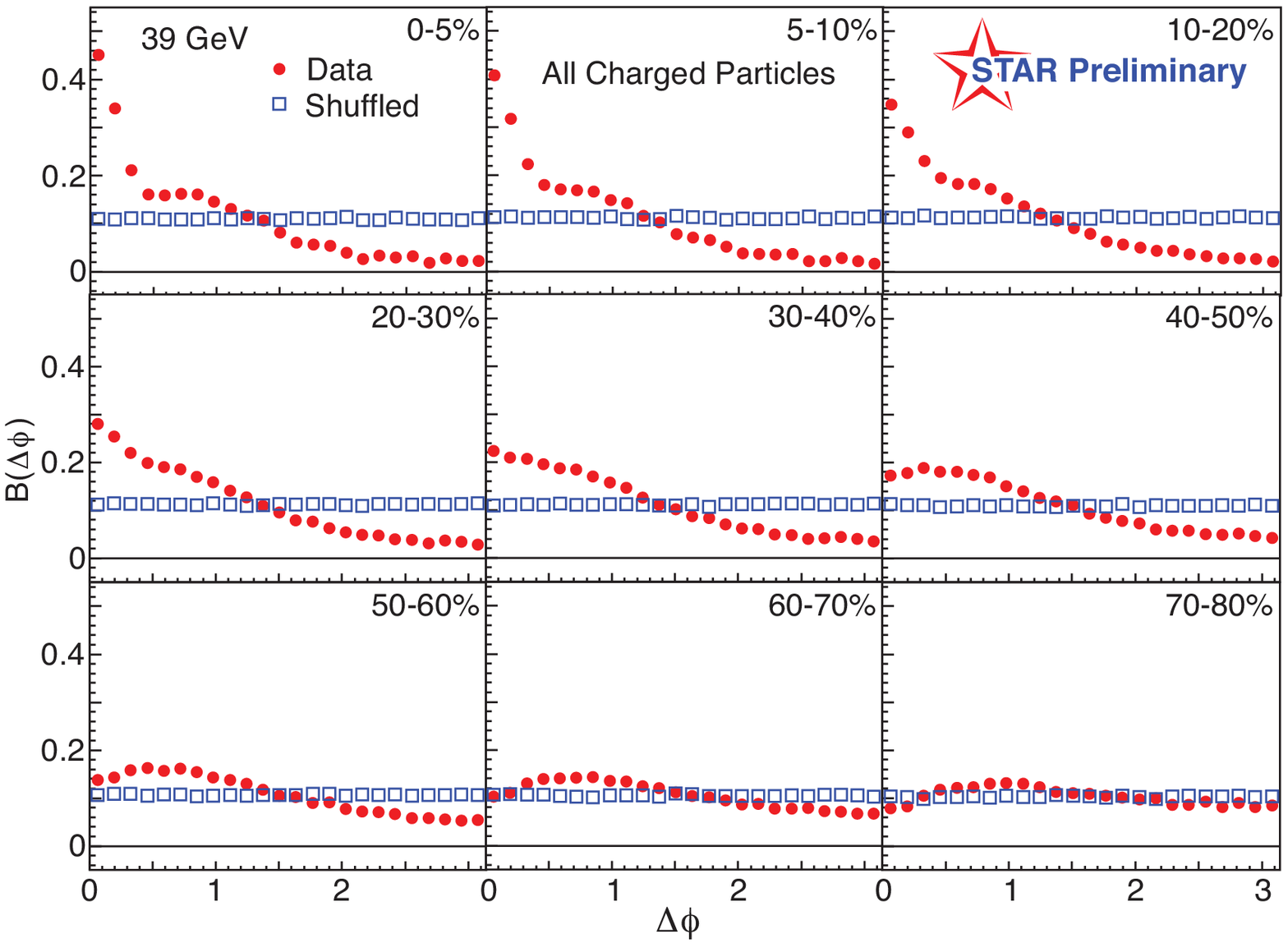}
\caption{\label{fig:fig05}The balance function in terms of  $\Delta \phi$ for all charged particle pairs from Au + Au collisions at $\sqrt{s_{\rm NN}}$  = 39 GeV for nine centrality bins.}
\end{minipage}\hspace{1pc}%
\begin{minipage}[t]{19pc}
\includegraphics[width=19pc]{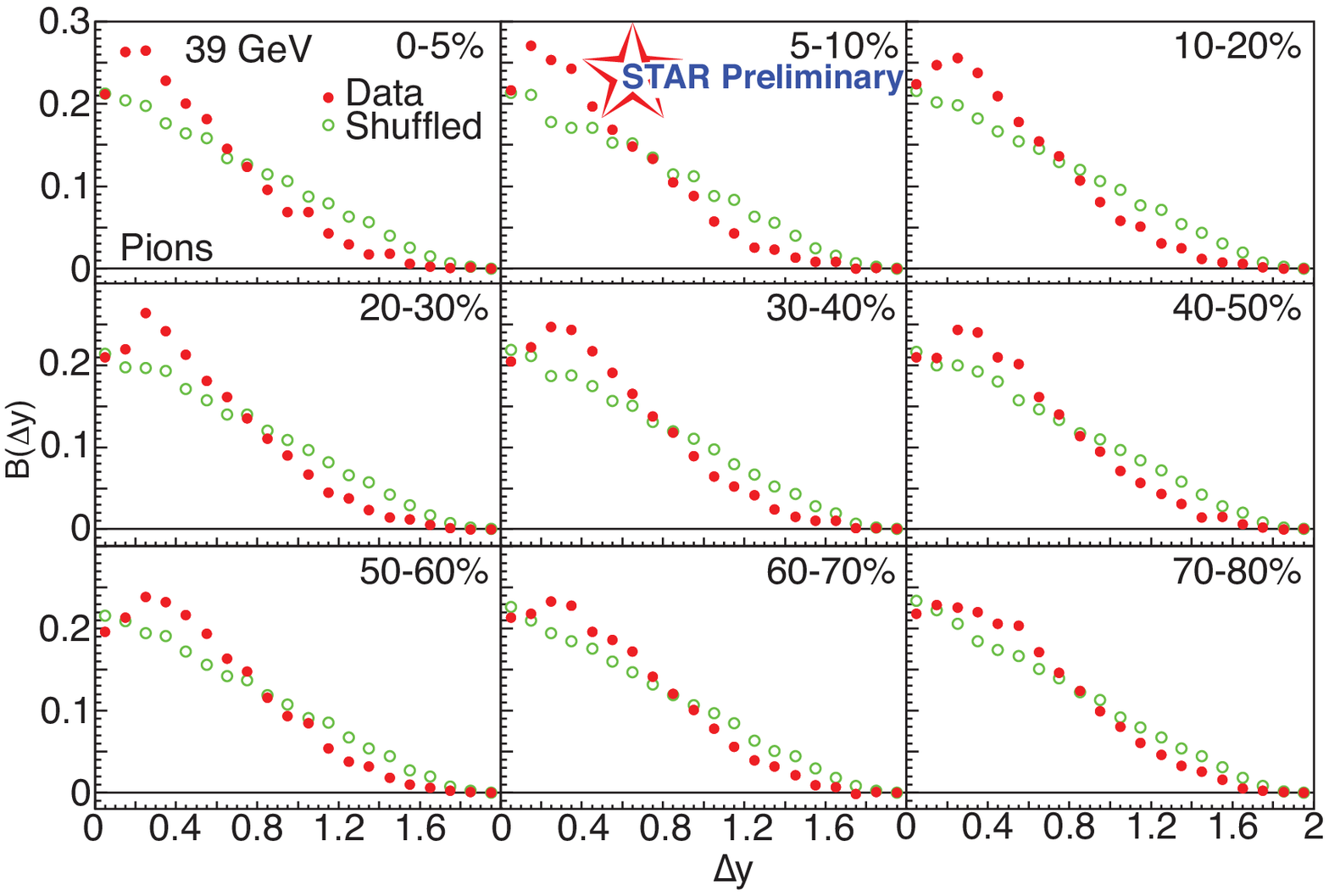}
\caption{\label{fig:fig06}(Color online) The balance function in terms of  $\Delta y$ for identfied pion pairs from Au + Au collisions at $\sqrt{s_{\rm NN}}$  = 39 GeV for nine centrality bins.}

\end{minipage} 

\end{figure}

The balance function in terms of  $ \Delta \phi$ would be more sensitive to transverse flow effect. Figure~\ref{fig:fig05} shows the balance function in terms of $\Delta \phi $ for all charged particles for nine centrality bins. The data shows a peak at around $\Delta \phi = 0$ in central collision while it is almost flat in peripheral collisions. This is due to strong radial flow in central collision, when charge pairs are created, they are strongly correlated in space and time due to conservation laws, if a flow of perfect liquid established in the system, charges would remain correlated throughout the whole process which would provide this observed peak at $\Delta \phi = 0$. Also due to STAR's uniform $\eta$ acceptance, the balance function for shuffled events  are always flat for all centralities.

Study of identified particle's balance function could reveal the charge production mechanisms  of different particle species. Figure~\ref{fig:fig06} show the balance functions for identified charged-pion pairs for nine centrality bins in terms of the relative rapidity. The balance function for experimental data gets narrower in central collisions, while balance functions for shuffled events are much more wider. The dip of balance function for $\Delta y < 0.2$ should due to interpair correlations (HBT, etc) according to model predictions \cite{Balance_HBT}.

\subsection{Beam Energy Dependence of the Balance Function Widths}

To quantify the correlation measured from balance function, we calculate the  weighted averages of $<\Delta \eta>$, $<\Delta \phi>$, and $<\Delta y>$ using Eq.(\ref{eq:weight}). To remove the interpair correlation, the weighted average is calculated within $ 0.1 \le \Delta\eta  \le 2.0 $ for $<\Delta \eta>$ and  $ 0.2 \le \Delta y  \le 2.0 $ for $<\Delta y>$

\begin{eqnarray}
\label{eq:weight}
\fl
 < \Delta \eta  >  = \frac{{\Sigma _i B(\Delta \eta _i )\Delta \eta _i }}{{\Sigma _i B(\Delta \eta _i )}}.
\end{eqnarray}

\begin{figure}[h]
\includegraphics[width=36pc]{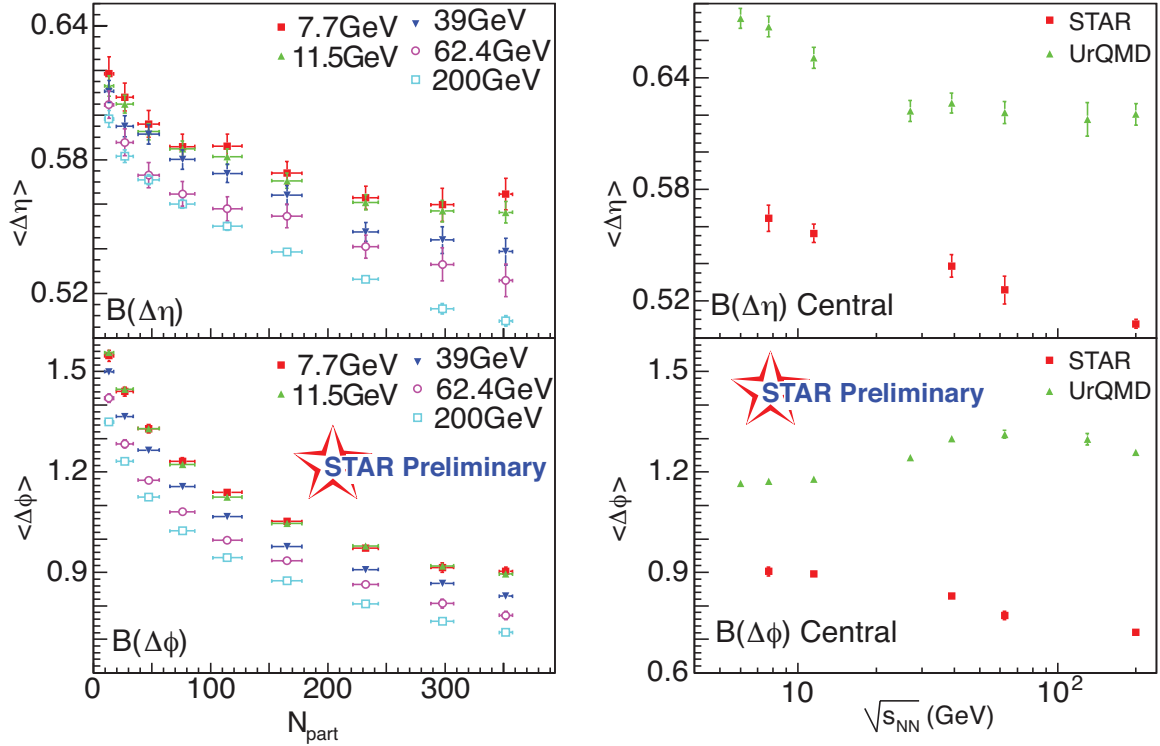}
\caption{\label{fig:fig07}The balance function widths in terms of  $\Delta \eta$, $\Delta \phi$ for all charged particle pairs from Au + Au collisions. Left panel shows centrality dependence, while the right panel shows beam energy dependence of most central evens (0-5\%). }

\end{figure}

The left panel of Figure~\ref{fig:fig07} shows the balance function widths for all charged particle in terms of the number of participating nucleons at $\sqrt{s_{\rm NN}}$ = 200 , 62.4, 39, 11.5, and 7.7 GeV. For both $\Delta \eta$ and $\Delta \phi$, the balance function widths narrow in central collisions for all energies presented here. The upper right panel of Figure~\ref{fig:fig07} shows the energy dependence of the balance function widths of $\Delta \eta$ for most central events(0-5\%). The observed $<\Delta \eta>$ from data shows a smooth decrease with increasing beam energy, which is consistent with the effect that radial flow is stronger at higher collision energies.  A UrQMD \cite{UrQMD} model calculation incorporating STAR acceptance filter show the similar trend but over predict the value, which is due to early hadronization time and less flow in UrQMD. An early hadronization time would cause a wider balance function, while relative diffusion and  interaction after freeze-out could further broaden it. The lower right panel of Figure~\ref{fig:fig07} shows the same thing for $<\Delta \phi>$. The experimental data shows a decreasing balance function widths with increasing beam energy, while UrQMD's width shows a little increase with increasing beam energy. This could due to final state interaction of produced particles in the UrQMD model.

\begin{figure}[h]

\includegraphics[width=36pc]{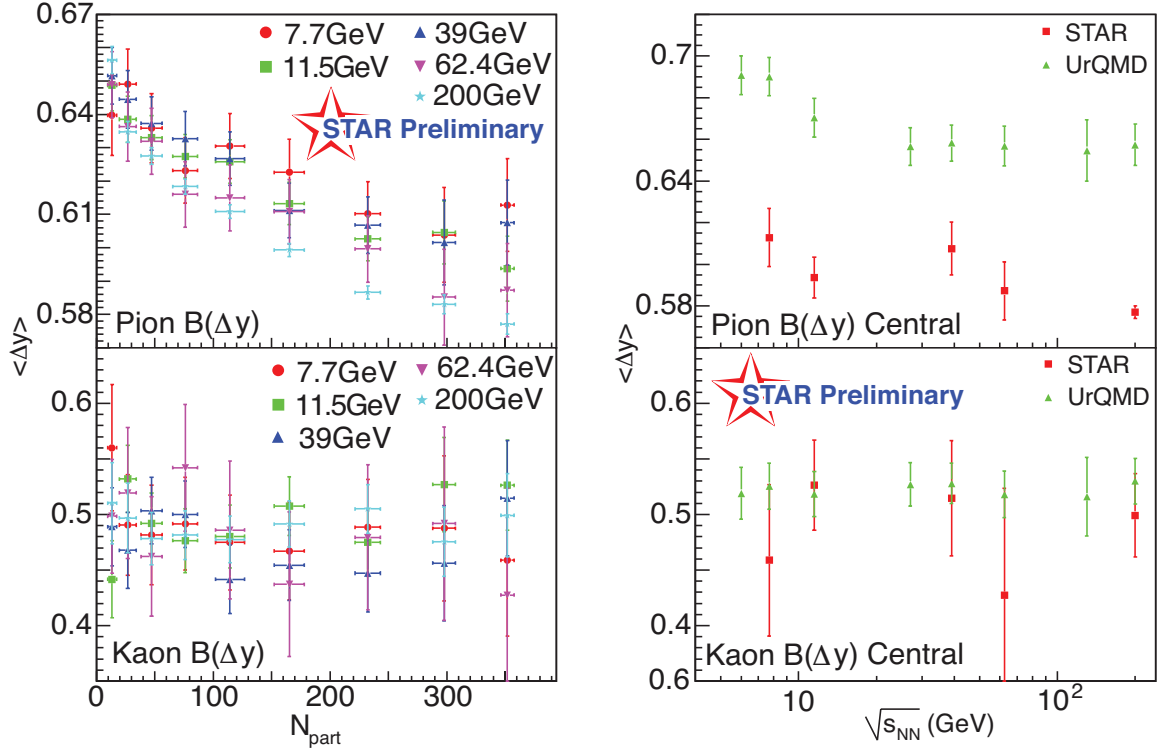}
\caption{\label{fig:fig08}(Color online) The balance function widths in terms of  $\Delta y$ for identified pions and kaons from Au + Au collisions. Left panel shows centrality dependence, while the right panel shows beam energy dependence of most central evens (0-5\%). }

\end{figure}

Figure~\ref{fig:fig08} shows widths of the balance function for identified charged pions and identified charged kaons for both centrality dependence and beam energy dependence. For identified pions, the measured balance function widths get narrower in both central collisions and higher energies, the UrQMD model shows a similar energy dependence but over predicts the signal. In contrast, the widths of the measured balance function for identified kaons show little centrality or beam energy dependence, while it is consistent with the predictions of UrQMD model. This lack of centrality and beam energy dependence could indicate the different production mechanism of kaons: they are produced mainly at the beginning of the collision rather than during a later hadronization stage. However, the balance function for kaons is also affected by the decay $\phi  \to K^ +   + K^ -  $, a further study of balance function for $q_{inv}$ could help to distinguish the contribution from $\phi$ decay.

\section{Summary}
In this paper, we have presented new results from the reaction-plane-dependent balance function, the reaction-plane-dependent balance function analysis gives the same difference between the like-sign and unlike-sign charge dependent azimuthal correlations as the three point correlator results published by STAR. A thermal blast-wave model incorporating local charge conservation and flow can reproduce most of the difference between like- and unlike-sign charge-dependent azimuthal correlation. 

We also measured the none event-plane-dependent balance function for Au+Au collisions at $\sqrt{s_{\rm NN}}$ = 200 , 62.4, 39, 11.5, and 7.7 GeV for all charged particles, identified charged pions, and identified charged kaons. We find that the balance functions in terms of $\Delta \eta$ and $\Delta \phi$ for all charged particles narrow both in central collisions
and higher collision energies, which is consistent with the delayed hadronization picture. The UrQMD model reproduced the trend for energy dependence, but over predicted the value of balance function widths. For identified kaons, the balance function widths show little centrality or beam energy dependence, which indicate that kaons might be created early in the collision. However, a detail study using balance function for $q_{inv}$ is necessary to remove the effect from $\phi$ decay.

\section{Acknowledgements}

We thank S. Schlichting, S Pratt for enlightening discussions.  We thank the RHIC Operations Group and RCF at BNL, the NERSC Center at LBNL, and the Open Science Grid consortium for providing resources and support. This work was supported in part by the Offices of NP and HEP within the US Department of Energy Office of Science, the US NSF, the Sloan Foundation; the DFG cluster of excellence ÒOrigin and Structure of the UniverseÓ of Germany; CNRS/IN2P3, STFC, and EPSRC of the United Kingdom; FAPESP CNPq of Brazil; Ministry of Education and Science of the Russian Federation; NNSFC, CAS, MoST, and MoE of China; GA and MSMT of the Czech Republic; FOM and NWO of the Netherlands; DAE, DST, and CSIR of India; Polish Ministry of Science and Higher Education; Korea Research Foundation; Ministry of Science, Education, and Sports of the Republic of Croatia; and the Russian Ministry of Science and Technology and RosAtom of Russia.

\end{document}